\documentclass[preprint,showpacs,preprintnumbers,amsmath,amssymb]{revtex4}
\def\bc{\begin{center}}

\def\ec{\end{center}}
\def\be{\begin{eqnarray}}
\def\ee{\end{eqnarray}}

\newcommand{\omits}[1]{}

\usepackage{amssymb}
\usepackage{amscd}
\usepackage{amsthm}
\usepackage{color}

 \omits{\textheight 24.5cm \textwidth 18.0cm
\oddsidemargin -1.5cm     
\topmargin -1.5cm}

\definecolor{dyellow}{rgb}{1.,0.8,.0}
\definecolor{myblue}{rgb}{.1,.1,.7}
\definecolor{dcyan}{rgb}{.0,.6,.6}
\definecolor{dmagenta}{rgb}{0.6,0.0,0.6}
\definecolor{brown}{rgb}{0.6,0.2,0.}
\definecolor{darkblue}{rgb}{.0,.0,0.5}
\definecolor{darkred}{rgb}{0.75,0.0,0.0}
\definecolor{orange}{rgb}{1.,.6,.0}
\definecolor{dorange}{rgb}{0.8,.4,.0}
\definecolor{darkgreen}{rgb}{0.0,0.6,0.0}
\definecolor{purple}{rgb}{.4,.0,.4}

\def\red{\color{red}}

\newcommand{\comments}[1]{}
\newcommand{\deff}{\emph}
\newcommand{\R}{\mathbb{R}}

\newcommand{\OmegaEL}{\Omega_{\mathrm{EL}}}

\newcommand{\HEL}{H_{\mathrm{EL}}}
\newcommand{\HdR}{H_{\mathrm{dR}}}
\newcommand{\XS}{\mathcal{X}_{\mathrm{S}}}
\newcommand{\XH}{\mathcal{X}_{\mathrm{H}}}

\newcommand{\vect}{\boldgrk}
\newcommand{\dd}{\mathrm{d}}
\newcommand{\tr}{\mathrm{tr\,}}
\newcommand{\im}{\mathrm{im}}
\newcommand{\Lied}[1]{\mathcal{L}_{\vect{#1}}}

\newcommand{\h}{\hat{h}}
\newcommand{\e}{\hat{e}}
\newcommand{\f}{\hat{f}}

\newcommand{\Span}{\mathrm{span}}
\omits{
\newcommand{\dd}{\mathrm{d}}

\newcommand{\vect}{\mathbf}
\newcommand{\Lied}[1]{\mathcal{L}_{\vect{#1}}}
\newcommand{\OmegaEL}{\Omega^1_{\mathrm{EL}}}

\newcommand{\XS}{\mathcal{X}_{\mathrm{S}}}
\newcommand{\XH}{\mathcal{X}_{\mathrm{H}}}
\newcommand{\tr}{\mathrm{tr\,}}
\newcommand{\HEL}{H_{\mathrm{EL}}}
\newcommand{\HdR}{H_{\mathrm{dR}}}

\newcommand{\h}{\hat{h}}
\newcommand{\e}{\hat{e}}
\newcommand{\f}{\hat{f}}

\newcommand{\Span}{\mathrm{span}}

\newcommand{\comments}[1]{{}}
\newcommand{\tedious}[1]{}
}

\newtheorem{thm}{Theorem}
\newtheorem{corol}{Corollary}[thm]
\newtheorem{lem}{Lemma}
\newtheorem{proposition}{Proposition}

\newcounter{defc}
\omits{\newenvironment{definition}{\stepcounter{defc}
\textbf{Definition \arabic{defc}.}}
  {\newline}
}
\newenvironment{definition}{\textbf{Definition:\ }}
  {\newline}

\newcounter{remkc}

\begin{document}


\title{The Generalized Liouville's Theorems via Euler-Lagrange Cohomology
  Groups on Symplectic Manifold}

\author{{Han-Ying Guo}$^{1,2}$} \email{hyguo@itp.ac.cn}
\author{{Jianzhong Pan}$^{3}$}
\email{pjz@mail.amss.ac.cn}
\author{{Bin Zhou}$^{4}$}
\email{zhoub@itp.ac.cn} \omits{\author{${D}^{1,2}$}
\email{}}
\affiliation{%
${}^1$ CCAST (World Laboratory), P.O. Box 8730, Beijing
   100080, China,}

\affiliation{%
${}^2$ Institute of Theoretical Physics,
 Chinese Academy of Sciences,
 P.O.Box 2735, Beijing 100080, China.}

 \affiliation{%
${}^3$Institute of Mathematics, Academy of Mathematics and System
Science, Chinese Academy of Sciences,  Beijing
   100080, China,}

\affiliation{%
${}^4$ Interdisciplinary Center of Theoretical Studies,
  Chinese Academy of Sciences, P.O. Box 2735, Beijing 100080, China.}

\date{August, 2004}

\begin{abstract}

Based on the Euler-Lagrange cohomology groups
$H_{EL}^{(2k-1)}({\cal M}^{2n}) (1 \leqslant k\leqslant n)$ on
symplectic manifold $({\cal M}^{2n}, \omega)$, their properties
and a kind of classification of vector fields on the manifold, we
generalize Liouville's theorem in classical mechanics to two
sequences, the symplectic(-like) and the Hamiltonian-(like)
Liouville's theorems. This also generalizes Noether's theorem,
since the sequence of symplectic(-like) Liouville's theorems link
to the cohomology directly.
\end{abstract}

\pacs{
45.20.Jj, 
02.40.Re, 
02.90.+p, 
05.45.-a, 
02.30.Hq 
}

\keywords{Symplectic manifold, Euler-Lagrange cohomology,
Liouville's theorem, Noether's theorem, volume-preserving system}

\maketitle

\tableofcontents

\section{Introduction}

It is well known that the theory on symplectic manifolds plays an
important role in classical mechanics (see, for example,
\cite{Arnold,AM}). And both Lagrangian and Hamiltonian mechanics, which
describe mechanical systems with potentials, have been already
well established. However, there are  many dynamical systems,
which have no  potentials, such as general volume preserving
systems on the symplectic manifolds. How to characterize and
describe these systems is an important problem. In addition,
the famous Liouville's theorem, which claims that the phase flow of a
Hamiltonian system preserves the phase volume,   also plays an
important role not only in classical mechanics but also in
physics. But, what about those general volume preserving systems
without potential on symplectic manifold? This is another
important problem.

Recently, the Euler-Lagrange cohomology has first been introduced
and discussed in \cite{ELcoh2,ELcoh4} for classical mechanics and
field theory in order to explore the relevant topics in the
(independent variable(s)) discrete mechanics and field theory
including symplectic and multisymplectic algorithms
\cite{ELcoh2,ELcoh4,gw}. Based upon these work, we have further
found that there is, in fact, a sequence of cohomology groups,
called the Euler-Lagrange cohomology groups, on the symplectic
manifolds \cite{zgpw}, \cite{zgw}, \cite{gpwz1}. What has been
found in \cite{ELcoh2, ELcoh4, gw} for the classical mechanics
with potential is the first one, constructed via the
Euler-Lagrange $1$-forms. We have also found that these cohomology
groups may play some important roles in the classical mechanics
with and without potential as well as other dynamical systems on
the symplectic manifolds, such as the volume-preserving systems
and so on \cite{zgpw, zgw, gpwz1}.

In this paper, we show that based on these Euler-Lagrange
cohomology groups and relevant issues the famous Liouville's
theorem in classical mechanics should be generalized to two
sequences of theorems: the sequence of symplectic(-like)
Liouville's theorems and the  Hamiltonian(-like) sequence. For the
symplectic(-like) sequence, it does not require the equations of
motion of the specified mechanical system hold on the symplectic
manifold as the phase space of the system. While for the
Hamiltonian(-like) sequence, it does always require on the
solutions of the corresponding mechanical system, similar to the
famous Liouville's theorem requires on the phase flow of a given
Hamiltonian system. In fact, the famous Liouville's theorem is the
first one in the Hamiltonian(-like) sequence and corresponds to
the image of the first Euler-Lagrange group. Further, the
generalization of Liouville's theorem implies a kind of
generalization of Noether's theorem.

In order to be self-contained, we first systematically introduce
the general definition of these Euler-Lagrange cohomology groups
$\HEL^{(2k-1)}({\cal M}^{2n},\omega)$, with $1 \leqslant k
\leqslant n$, on a $2n$-dimensional symplectic manifold $({\cal
M}^{2n},\omega)$ and study their properties in some details. In
fact, for each $k \leqslant n$, the Euler-Lagrange cohomology
group $\HEL^{(2k-1)}(\mathcal{M},\omega)$ is a quotient group of
the $(2k-1)$st symplectic(-like) vector fields over the $(2k-1)$st
Hamiltonian(-like) ones (see  the definitions in next section).
Thus, these cohomology groups classify the vector fields on
symplectic manifold. We show that, for $k = 1$ and $k = n$, they
are isomorphic to the corresponding de~Rham cohomology groups
$H_{\rm d R}^{(1)}({\cal M}^{2n})$ and $H_{\rm d R}^{(2n-1)}({\cal
M}^{2n})$, respectively. Consequently, due to the Poincar\'e
duality, the first Euler-Lagrange cohomology group $H_{\rm
EL}^{(1)}({\cal M}^{2n},\omega)$ and the highest one $H_{\rm
EL}^{(2n-1)}({\cal M}^{2n},\omega)$ are dual to each other if
$\mathcal{M}^{2n}$ is closed. We also show that the other
Euler-Lagrange cohomology groups $H_{\rm EL}^{(2k-1)}({\cal
M}^{2n},\omega)$ for $1<k<n$ are different from either the de~Rham
cohomology groups or the harmonic cohomology groups on $({\cal
M}^{2n},\omega)$, in general.\omits{ Some aspects on these issues
have been briefly given in \cite{zgw}.}

{From} the cohomological point of view, the ordinary Hamiltonian canonical
equations correspond to 1-forms that represent the trivial element in the
first Euler-Lagrange group $\HEL^{(1)}({\cal M}^{2n},\omega)$.
Analogous with this fact, it is natural and significant to find the general
volume-preserving equations on $({\cal M}^{2n},\omega)$ from such forms that
represent the trivial element in the highest Euler-Lagrange cohomology group
$H_{\rm EL}^{(2n-1)}({\cal M}^{2n},\omega)$. 
We have introduced this general kind of volume-preserving
equations from this point of view. In general, there are no
potentials for these volume-preserving systems described by the
equations. Only for the special cases, these equations become the
ordinary canonical equations in the Hamilton mechanics. Therefore,
the Hamilton mechanics has been generalized to the
volume-preserving systems on symplectic manifolds via the
cohomology  \cite{zgpw, zgw, gpwz1}.

In classical mechanics, for a Hamiltonian system on 
$(\mathcal{M}^{2n},\omega)$, there is a class of phase-area conservation laws
for its phase flow $g^t$. Conservations of both the symplectic 2-form $\omega$
and the volume form $\tau = \frac{1}{n!}\,\omega^n$ on $\cal M$
are included (see, for example, \cite{Arnold}):%
 \be\label{areapreserv'} \nonumber\int_{g^t
\sigma^2}\omega &=& \int_{\sigma^2}\omega, ~\forall 2{\rm -chain}~
\sigma^2 \subset
{\mathcal{M}^{2n}},\\\nonumber%
&&\vdots\\
 \int_{g^t \sigma^{2k}}\omega^k &=&
\int_{\sigma^{2k}}\omega^k, ~\forall 2k{\rm-chain}~ \sigma^{2k}
\subset {\mathcal{M}^{2n}},\\\nonumber%
 && \vdots\\\nonumber
 \int_{g^t \sigma^{2n}}\omega^{n} &=& \int_{ \sigma^{2n}}\omega^{n},
~\forall 2n{\rm -chain} ~\sigma^{2n} \subset
{\mathcal{M}^{2n}},\\\nonumber \ee%
 where the power of $\omega$  is
in the wedge product. The last one is the famous Liouville's
theorem. \omits{}Based on the properties of the cohomological
classification of the vector spaces, we generalize the Liouville's
theorem. The original Liouville's theorem with respect to the
phase flows may be called the Hamiltonian Liouville's theorem. It
can be generalized first to the symplectic Liouville's theorem
requiring the system to move along a symplectic flow generated by
a symplectic vector field.
Then we further generalize these two Liouville's theorems  to the
ones with respect to the $(2k-1)$st Euler-Lagrange group
$H_{EL}^{(2k-1)}(\mathcal{M},\omega)$, $1<k \leqslant n$,
\omits{i.e. the quotient space of
the $2k-1$st symplectic(-like) vector fields over the $2k-1$st
Hamiltonian(-like) ones. Thus, Liouville's theorem may be
generalized to}namely the $(2k-1)$st symplectic(-like) and
Hamiltonian(-like) Liouville's theorem, respectively. For the case
of $k=n$, the highest Euler-Lagrange cohomology group,
 it gives the
most general symplectic(-like) and Hamiltonian(-like) Liouville's
theorems\omits{ according to the kernel and image of this highest
cohomology}, respectively.

As was just mentioned,  this generalization of Liouville's theorem
directly leads to a kind of generalization of the famous Noether's
theorem for the conservation laws via symmetries. As far as the
classical mechanical systems are concerned, all known conservation
laws are always associated with certain symmetries and hold on the
solution space of the equations of motion of the system.  As a
matter of fact, corresponding to the kernel and image of the
$(2k-1)$st Euler-Lagrange cohomology group, there is $(2k-1)$st
degree symplectic(-like) and Hamiltonian(-like) area conservation
laws, respectively. The former requires the closeness condition of
the Euler-Lagrange $(2k-1)$-forms or the vector fields being
$(2k-1)$st
symplectic(-like). %
While, the latter
requires the exactness of the Euler-Lagrange $(2k-1)$-forms or the
vector fields be $(2k-1)$st Hamiltonian(-like). %
In other words, among two kinds of conservation laws for the area
preserving of the $2k$-dimensional chain $\sigma^{2k} \subset
{\cal M}$, the symplectic(-like) ones\omits{} generalizes
Noether's theorem via the Euler-Lagrange cohomology groups.

This paper is arranged as follows. In section 2, we first briefly
recall the definition of the first Euler-Lagrange cohomology group
on a symplectic manifold $({\cal M}^{2n},\omega)$ and prove that
it is isomorphic to the first de~Rham cohomology group on the
manifold. Then we introduce the general definition of the
$(2k-1)$st Euler-Lagrange cohomology groups for $1\leqslant k
\leqslant n$ on $({\cal M}^{2n},\omega)$ and indicate that the
highest one is equivalent to the $(2n-1)$st de~Rham cohomology
group. We also indicate that in general they are not isomorphic to
each other and that they are not isomorphic to either the de~Rham
cohomology or the harmonic cohomology on $({\cal M}^{2n},\omega)$.
The relative Euler-Lagrange cohomology is also introduced in
analog with the relative de~Rham cohomology. The general
volume-preserving equations are introduced in section 3. Their
relations with ordinary canonical equations in the Hamilton
mechanics as well as other volume-preserving systems are
discussed. It is clear that the general volume-preserving
equations are the generalization of the ordinary canonical
equations in Hamilton Mechanics. \omits{\cite{gpwz}, \cite{zgpw},
\cite{zgw},
 \cite{gpwz1}. }In the section 4, based on these results, we 
generalize Liouville's theorem in mechanics \omits{ that require
the systems satisfying the canonical equations }to the generalized
$(2k-1)$st symplectic(-like) and Hamiltonian(-like) ones, respectively.
Finally, we end with some conclusion and remarks.

\section{The Euler-Lagrange Cohomology Groups on Symplectic Manifold}

\subsection{The Hamilton Mechanics, First and Higher Euler-Lagrange Cohomology}
As is well known, the ordinary canonical equations  can be
expressed as
\begin{equation}\label{cnleqs}
  - i_{\vect{X}_H}\omega = \dd H,
\end{equation}
where $\vect{X}_H$ denotes the Hamiltonian vector field with
respect to the Hamiltonian $H$. Introducing what is called the
Euler-Lagrange 1-form
\begin{eqnarray}\label{el1form}
  E^{(1)}_{\vect{X}}:=- i_{\vect{X}}\omega, 
\end{eqnarray}
where ${\vect{X}}$ is an arbitrary vector field of degree one, the
eq.~(\ref{cnleqs}) indicates that $E^{(1)}_{\vect{X}_H}$ is exact.
In other words, it belongs to the image part of the (first)
Euler-Lagrange cohomology group \cite{gw}
\begin{equation}\label{elc}
  \nonumber
  H^{(1)}_{\rm EL}({\cal M},\omega):=Z^{(1)}_{\rm EL}({\cal M},\omega)/
  B^{(1)}_{\rm EL}({\cal M},\omega),
\end{equation}
where
\begin{eqnarray*}
  Z^{(1)}_{\rm EL}(\mathcal{M},\omega)
  & := & \{E^{(1)}_{\vect{X}}|\,\dd E^{(1)}_{\vect{X}}=0 \}
  = \ker(\dd)\cap\Omega^1(\mathcal{M}),
  \\
  B^{(1)}_{\rm EL}(\mathcal{M},\omega)
  & := & \{E^{(1)}_{\vect{X}}|\, E^{(1)}_{\vect{X}}=\dd \beta \}
  = \im(\dd)\cap\Omega^1(\mathcal{M}).
\end{eqnarray*}
On the other hand, the vector fields corresponding to the kernel part
are (degree one) symplectic by definition (See, e.g.,
\cite{lecture}). Therefore, the cohomology can also equivalently
be defined as
\begin{equation}
  \label{elc1}\nonumber
  H^{(1)}_{\rm EL}(\mathcal{M},\omega)
  := \XS^{(1)}(\mathcal{M},\omega)/ \XH^{(1)}(\mathcal{M},\omega),
\end{equation}
due to 
$
  Z^{(1)}_{\rm EL}(\mathcal{M},\omega) \cong \XS^{(1)}(\mathcal{M},\omega), 
$
$
  B^{(1)}_{\rm EL}(\mathcal{M},\omega) \cong
  \XH^{(1)}(\mathcal{M},\omega),
$ 
where the degree one of the vector fields indicates that they
correspond to the Euler-Lagrange 1-forms.

In general, for  $1\leqslant k \leqslant n$, we may define the
Euler-Lagrange $(2k-1)$-forms
\begin{eqnarray}\label{el2k-1form}
  E^{(2k-1)}_{\vect{X}}:=- i_{\vect{X}}\omega^k, 
\end{eqnarray}
and define the sets of (degree $2k-1$) symplectic(-like) and Hamiltonian(-like)
vector fields, respectively, as
\begin{eqnarray}\label{V}
  \XS^{(2k-1)}(\mathcal{M},\omega) & := & \{\,\vect{X}\in\mathcal{X(M)}\,|\,
   \dd E^{(2k-1)}_{\vect{X}}= 0 \,\}, \\\nonumber
  \XH^{(2k-1)}(\mathcal{M},\omega) & := & \{\,\vect{X}\in\mathcal{X(M)}\,|\,
  E^{(2k-1)}_{\vect{X}} \textrm{ is exact}\}.
\end{eqnarray}
Here\omits{ upper index ${(2k-1)}$ denotes they corresponding to
the Euler-Lagrange ${(2k-1)}$-forms,} $\mathcal{X(M)}$ denotes the
space of vector fields on $\mathcal{M}$. It is easy to prove that
\begin{widetext}
\begin{eqnarray}
  &  \XS^{(1)}(\mathcal{M},\omega) \subseteq \ldots \subseteq
  \XS^{(2k-1)}(\mathcal{M},\omega)
  \subseteq \XS^{(2k+1)}(\mathcal{M},\omega)
  \subseteq \ldots \subseteq\XS^{(2n-1)}(\mathcal{M},\omega),&
\nonumber \\
  &  \XH^{(1)}(\mathcal{M},\omega) \subseteq \ldots \subseteq
  \XH^{(2k-1)}(\mathcal{M},\omega)
  \subseteq \XH^{(2k+1)}(\mathcal{M},\omega)
  \subseteq \ldots \subseteq\XH^{(2n-1)}(\mathcal{M},\omega),&
\label{eq:Hrels} \\
  &\XH^{(2k-1)}(\mathcal{M},\omega)
  \subseteq \XS^{(2k-1)}(\mathcal{M},\omega).&
\nonumber
\end{eqnarray}
\end{widetext}
In fact,  the (symplectic-like)  vector fields
$\XS^{(2k-1)}(\mathcal{M},\omega)$ is a Lie algebra under the
commutation bracket of vector fields, and
$\XH^{(2k-1)}(\mathcal{M},\omega)$ is an ideal of
$\XS^{(2k-1)}(\mathcal{M},\omega)$ because
\begin{equation}
  [\XS^{(2k-1)}(\mathcal{M},\omega),\XS^{(2k-1)}(\mathcal{M},\omega)]
  \subseteq \XH^{(2k-1)}(\mathcal{M},\omega).
\end{equation}
\omits{where $\XH^{(2k-1)}(\mathcal{M},\omega)$ is the Lie algebra
of the degree $(2k-1)$ Hamiltonian(-like) \omits{volume-preserving
}vector fields. }It is clear that, for $k = n$,
$\XS^{(2n-1)}(\mathcal{M},\omega)$ is the Lie algebra of the
(symplectic-like) volume-preserving vector fields, including all
other Lie algebras just mentioned  as its subalgebras.

The quotient Lie algebra
\begin{equation}\label{defH_EL}
  H^{(2k-1)}_{\mathrm{EL}}(\mathcal{M},\omega)
  := \XS^{(2k-1)}(\mathcal{M},\omega)/\XH^{(2k-1)}(\mathcal{M},\omega)
\end{equation}
is  called the {\it{$(2k-1)$st Euler-Lagrange cohomology group}}.
It is called and treated as a group because it is an Abelian Lie
algebra. \omits{Equivalently, it may also be defined in terms of
the kernel and the image of $\dd$ over the space of Euler-Lagrange
$(2k-1)$-forms corresponding to $\XS^{(2k-1)}(\mathcal{M},\omega)$
and $\XH^{(2k-1)}(\mathcal{M},\omega)$, respectively.

It can be proved that $H^{(1)}_{\mathrm{EL}}(\mathcal{M},\omega)$
and $H^{(2n-1)}_{\mathrm{EL}}(\mathcal{M},\omega)$ are linearly
isomorphic to the corresponding de~Rham cohomology groups
$H^{(1)}_{\mathrm{dR}}(\mathcal{M})$ and
$H^{(2n-1)}_{\mathrm{dR}}(\mathcal{M})$, respectively
\cite{gpwz,gpwz1}. In general, however, the Euler-Lagrange
cohomology $H^{(2k-1)}_{\mathrm{EL}}(\mathcal{M},\omega),~1< k <
n,$ is not isomorphic to either the de~Rham cohomology or the
harmonic cohomology on $\cal M$ \cite{gpwz,gpwz1}.}

On the other hand, for each $k$ ($1 \leqslant k \leqslant n$),
the Euler-Lagrange $(2k-1)$-forms $E_{\vect{X}}^{(2k-1)}$ as well
as the kernel and image spaces of them with respect to $\dd$ may
be introduced: \begin{eqnarray}
E_{\vect{X}}^{(2k-1)}(\cal M)&:=&-i_{\vect{X}}(\omega^k),
  ~~\vect{X}\in\mathcal{X}(\mathcal{M}, \omega);\\
  Z_{\rm EL}^{(2k-1)}(\mathcal{M},\omega)&:=&\{E_{\vect{X}}^{(2k-1)}|\
  \dd E_{\vect{X}}^{(2k-1)} = 0\};\\
  B_{\rm EL}^{(2k-1)}(\mathcal{M},\omega)
  &:=&\{E_{\vect{X}}^{(2k-1)}|\ E_{\vect{X}}^{(2k-1)}
  \textrm{ is exact}\}.
\end{eqnarray}
The \deff{$(2k-1)$st Euler-Lagrange cohomology group} may also be
equivalently defined as
\begin{equation}
  H^{(2k-1)}_{\mathrm{EL}}(\mathcal{M},\omega)
  :=Z_{\rm EL}^{(2k-1)}(\mathcal{M},\omega)
  /B_{\rm EL}^{(2k-1)}(\mathcal{M},\omega).
  \label{eq:HEL}
\end{equation}
The equivalence between (\ref{defH_EL}) and (\ref{eq:HEL}) is a
corollary of the following lemma:

\begin{lem}
  Let $x\in \mathcal{M}$ be an arbitrary point and
$\vect{X}\in T_x \mathcal{M}$. Then, for each
$1 \leqslant k \leqslant n$, $i_{\vect{X}}(\omega^k) = 0$ if and
only if $\vect{X} = 0$. \label{lem:inject}
\end{lem}
\omits{ 
\begin{proof}
We need only to prove that $i_{\vect{X}}(\omega^k) = 0$ implies
$\vect{X} = 0$. We assume that there is a nonzero vector
$\vect{X}\in T_x \mathcal{M}$ satisfying $i_{\vect{X}}(\omega^k) 0$ for some
$k$.
\end{proof}
}

\subsection{The Spaces $\XS^{(2k-1)}(\mathcal{M},\omega)$ and
  Highest Euler-Lagrange Cohomology\omits{
$H_{{\rm EL}}^{(2n-1)}(\mathcal{M},\omega)$}}\omits{\subsection{Some Operators}}
\label{sect:Operators}

In order to investigate the properties of the Euler-Lagrange
cohomology groups, it is needed to introduce some operators first.

For a point $x \in \mathcal{M}$, the symplectic form $\omega$ can
be locally expressed as the well-known formula $\omega = \dd
p_i\wedge\dd q^i$ in the Darboux coordinates $(q,p)$. Then let us
introduce a well defined linear map on $\Lambda^*_x(\mathcal{M})$:
\begin{equation}
  \f := i_{\frac{\partial}{\partial q^i}}i_{\frac{\partial}{\partial p_i}}.
\label{eq:f} \end{equation} Note that $\f = 0$ when acting on
$\Lambda^1(T^*_x \mathcal{M})$ or $\Lambda^0(T^*_x \mathcal{M})$.
And a map $\f$ can be defined on the exterior bundle $\Lambda^*(\mathcal{M})$.
Further, a linear homomorphism, denoted also by $\f$, can be obtained on
$\Omega^*(\mathcal{M})$, the space of differential forms. Especially, we have
the identity
\begin{equation}
  \f\omega = n.
\label{pomega}
\end{equation}

Another two operators
\begin{equation}
  \e: \Lambda^*_x(\mathcal{M}) \longrightarrow \Lambda^*_x(\mathcal{M}), \quad
  \alpha \longmapsto \e\alpha = \alpha\wedge\omega
\label{eq:e}
\end{equation}
and
\begin{equation}
  \h: \Lambda^k(T^*_x \mathcal{M})\longrightarrow \Lambda^k(T^*_x \mathcal{M}), \quad
  \alpha \longmapsto \h\alpha = (k-n)\,\alpha
\label{eq:h} \end{equation} can also be defined at each $x\in
\mathcal{M}$. \begin{lem}
  The operators $\e$, $\f$ and $\h$ on $\Lambda^*_x(\mathcal{M})$ satisfy
  \begin{equation}
    [\h,\e] = 2\,\e, \quad
    [\h,\f] = -2\,\f, \quad
    [\e,\f] = \h, \quad \forall\,x \,\in\, \cal M .
  \label{eq:efh}
  \end{equation}
\label{lem:efh}
\end{lem}
\begin{proof}
These relations can be verified directly. Here is a trickier
proof.

First we define some ``fermionic" operators on
$\Lambda^*_x(\mathcal{M})$
\begin{equation}
\begin{array}{lll}
  \psi_i := i_{\frac{\partial}{\partial q^i}}, & \qquad &
  \psi^i := i_{\frac{\partial}{\partial p_i}}, \\
  \chi_i : \alpha \longmapsto \dd p_i\wedge\alpha, & &
  \chi^i : \alpha \longmapsto \dd q^i\wedge\alpha.
\end{array} \label{eq:fermionic} \end{equation}
For these operators, it is easy to verify that the non-vanishing
anti-commutators  are \begin{equation}
  \{\psi_i,\chi^j\} = \delta^j_i, \qquad
  \{\psi^i,\chi_j\} = \delta^i_j.
\end{equation}
Given an integer $0 \leqslant k \leqslant 2n$, we can check that,
for any $\alpha\in\Lambda^k(T^*_x \mathcal{M})$,
$ 
  (\chi_i\,\psi^i + \chi^i\,\psi_i)\alpha = k\,\alpha.
$ 
Therefore,
\begin{equation}
  \h = \chi_i\,\psi^i + \chi^i\,\psi_i - n.
\end{equation}
According to the definitions,
\begin{equation}
  \e = \chi_i\,\chi^i, \qquad \f = \psi_i\,\psi^i.
\end{equation}
Then the relations in eqs.~(\ref{eq:efh}) can be obtained when
$\e$, $\f$ and $\h$ are viewed as bosonic operators.
\end{proof}

\comments{
\begin{corol}
  Let $x \in M$ be an arbitrary point and $0 \leqslant k \leqslant 2n$.
  If $\alpha \in \Lambda^k(T^*_x M)$ satisfies $\alpha\wedge\omega = 0$, then
  $(\f\alpha)\wedge\omega = - (n-k)\, \alpha.$
\label{lem:kernelomega}
\end{corol}
}

For a point $x \in \mathcal{M}$ 
the following formulas can be derived recursively:
\begin{equation}
  [\e^k,\f] = k\,\e^{k-1}(\h + k - 1), \qquad
  [\e,\f^k] = k\,\f^{k-1}(\h - k + 1),
\label{eq:efl}
\end{equation}
where $k$ is an arbitrary positive integer. Then there is the lemma:

\begin{lem}
  Let $\alpha$ be a 2-form. If ~$\e^k \alpha=0$ for some $k<n-1$, then
$\alpha=0$. \label{lem:injective}
\end{lem}
\begin{proof}
  Applying both sides of the first equation in (\ref{eq:efl}) on $\alpha$,
  we have
  \begin{equation}
    \e^k \f \alpha = k \, (k-n+1)\, \e^{k-1} \alpha.
  \label{eq:wsy}
  \end{equation}
  Since $\f\alpha$ is a number at each point, the left hand side is
  $(\f\alpha)\, \omega^k$. Applying $\e$ on both sides, we get
  $$ (\f\alpha)\, \omega^{k+1}= k(k-n+1)\, \e^k \alpha=0. $$
  Since $k+1<n$, we have $\f\alpha=0$. Now the identity (\ref{eq:wsy}) becomes
  $$  k \, (k-n+1)\, \e^{k-1} \alpha=0. $$
  Since $k<n-1$, we obtain $\e^{k-1} \alpha=0.$ Therefore, the value of $k$ can be
  reduced by 1, and further it can be eventually reduced to 0.
\end{proof}

The above lemma implies that the map sending
$\alpha\in\Lambda^2(T^*_x \mathcal{M})$ to
$\alpha\wedge\omega^{n-2}\in\Lambda^{2n-2}(T^*_x \mathcal{M})$ is
an isomorphism.

\omits{\subsection{The Spaces $\XS^{(2k-1)}(\mathcal{M},\omega)$
and
  $H_{{\rm EL}}^{(2n-1)}(\mathcal{M},\omega)$}}

We have indicated in previous subsection that
$\XS(\mathcal{M},\omega) = \XS^{(1)}(\mathcal{M},\omega) \subseteq
\XS^{(2k-1)}(\mathcal{M},\omega)$ for each possible $k$. The
following theorem tells us much more.
\begin{thm}
  Let $(\mathcal{M},\omega)$ be a $2n$-dimensional symplectic manifold with
$n \geqslant 2$. Then, for each $ k \in \{1, 2, \ldots,  n - 1
\}$,
  \begin{equation}
  \XS^{(2k-1)}(\mathcal{M},\omega) = \XS(\mathcal{M},\omega).
  \end{equation}
\label{thm:XS}
\end{thm}
\begin{proof} We need only to prove that
$\XS^{(2k-1)}(\mathcal{M},\omega) \subseteq
\XS(\mathcal{M},\omega)$ for each $k\in \{1, 2, \dots, n-1\}$.

In fact, for any $\vect{X}\in\XS^{(2k-1)}(\mathcal{M},\omega)$,
\begin{displaymath}
  \Lied{X}(\omega^k) = k\,(\Lied{X}\omega)\wedge\omega^{k-1} = 0.
\end{displaymath} Since $0 \leqslant k-1
\leqslant n-2$ while $\Lied{X}\omega$ is a 2-form, we can use
Lemma \ref{lem:injective} pointwisely. This yields $\Lied{X}\omega
=0$. Thus, $\vect{X}\in\XS(\mathcal{M},\omega)$. This proves
$\XS^{(2k-1)}(\mathcal{M},\omega)\subseteq\XS(\mathcal{M},\omega)$
when $1 \leqslant k \leqslant n-1$.
\end{proof}
\begin{corol}
For each $k < n$, $[\XS^{(2k-1)}(\mathcal{M},\omega),
\XS^{(2k-1)}(\mathcal{M},\omega)]\subseteq\XH(\mathcal{M},\omega).$
\end{corol}

As was implied by Lemma \ref{lem:inject}, the map
$\mathcal{X}(\mathcal{M})\longrightarrow\Omega^{2n-1}(\mathcal{M}),
\vect{X}\longmapsto i_{\vect{X}}(\omega^n)$ is a linear
isomorphism. From $\Lied{X}(\omega^n) = \dd
i_{\vect{X}}(\omega^n)$, it follows that
$\XS^{(2n-1)}(\mathcal{M},\omega)$ is isomorphic to
$Z^{(2n-1)}(\mathcal{M})$, the space of closed $(2n-1)$-forms.
Lemma \ref{lem:inject} also implies that
$\XH^{(2n-1)}(\mathcal{M},\omega)$ is isomorphic to
$B^{(2n-1)}(\mathcal{M})$, the space of exact $(2n-1)$-forms.
These can be summarized as in the following theorem:
\begin{thm} The
linear map $\mu_n: \mathcal{X}(\mathcal{M})\longrightarrow
\Omega^{2n-1}(\mathcal{M}),\vect{X} \longmapsto
i_{\vect{X}}(\omega^n)$ is an isomorphism. Under this isomorphism,
$\XS^{(2n-1)}(\mathcal{M},\omega)$ and
$\XH^{(2n-1)}(\mathcal{M},\omega)$ are isomorphic to
$Z^{(2n-1)}(\mathcal{M})$ and $B^{(2n-1)}(\mathcal{M})$,
respectively. \label{thm:thm6}
\end{thm}

\begin{corol}
  The $(2n-1)$st Euler-Lagrange cohomology group
$H^{(2n-1)}_{\mathrm{EL}}(\mathcal{M},\omega)$ is linearly
isomorphic to $H^{(2n-1)}_{\mathrm{dR}}(\mathcal{M})$, the
$(2n-1)$st de~Rham cohomology group. \end{corol}

When $\mathcal{M}$ is closed, $\HEL^{(2n-1)}(\mathcal{M},\omega)$
is linearly isomorphic to the dual space of
$\HEL^{(1)}(\mathcal{M},\omega)$, because
  $\HdR^{(k)}(\mathcal{M}) \cong (\HdR^{(2n-k)}(\mathcal{M}))^*$
for such kind of manifolds. If $\mathcal{M}$ is not compact, this
relation cannot be assured.

\subsection{The Other Euler-Lagrange Cohomology Groups}
\label{sectRelation}

Although the first and the last Euler-Lagrange cohomology groups
can be identified with the corresponding de~Rham cohomology
groups, respectively, it is still valuable to know whether the
other Euler-Lagrange cohomology groups are nontrivial and
different from corresponding de~Rham cohomology groups in general.

In this subsection we will enumerate some examples and properties
relative to  this problem. We shall  point out that, for the torus
$T^{2n}$ with the standard symplectic structure $\omega$ and $n
\geqslant 3$, $\HEL^{(2k-1)}(T^{2n}, \omega)$ is not isomorphic to
$\HdR^{(2k-1)}(T^{2n})$ whenever $1 < k < n$ (see,
Corollary~\ref{cor:T2n}). In addition, we shall prove that there
is a $6$-dimensional symplectic manifold $({\cal M}^6,\omega)$ for
which the Euler-Lagrange cohomology group $\HEL^{(3)}
(\mathcal{M}^6,\omega)$ is not isomorphic to
$\HEL^{(1)}(\mathcal{M},\omega)$ (see, Theorem~\ref{thm:nilm}).
Therefore, these indicate that the Euler-Lagrange cohomology
groups other than the first and the last ones are some new
features of certain given symplectic manifolds.

Let $L_\alpha$ be the homomorphism defined by the cup product with
a cohomology class $[\alpha]$, where
 $\alpha$ is a representative. From the
definition, there is an injective homomorphism of vector spaces,
$$\pi_{2k-1}: H^{(2k-1)}_{\mathrm{EL}}(\mathcal{M},\omega)
\longrightarrow
  H^{(2k-1)}_{\mathrm{dR}}(\mathcal{M}), $$
for each $ k \in \{ 1, 2, \ldots, n \}$ such that the following
diagram is commutative:
\begin{widetext}
\begin{equation}
\begin{CD}
  H^{(1)}_{\mathrm{EL}}(\mathcal{M},\omega) @> >>
  H^{(3)}_{\mathrm{EL}}(\mathcal{M},\omega) @> >> \cdots
  @> >> H^{{(2n-3)}}_{\mathrm{EL}}(\mathcal{M},\omega)
\\
  @V \pi_1 VV   @V \pi_3 VV    @V \cdots VV    @V \pi_{2n-3} VV
\\
  H^{(1)}_{\mathrm{dR}}(\mathcal{M}) @> L_{\omega} >>
  H^{(3)}_{\mathrm{dR}}(\mathcal{M}) @> L_{\omega} >>
                \cdots @> L_{\omega} >> H^{(2n-3)}_{\mathrm{dR}}(\mathcal{M}).
\end{CD}
\label{eq:CD}
\end{equation}
\end{widetext}
In fact, for an equivalence class $[\vect{X}]_{(2k-1)} \in
H^{(2k-1)}_{\mathrm{EL}}(\mathcal{M},\omega)$ $(1 \leqslant k
\leqslant n)$ with $\vect{X} \in \XS^{(2k-1)}(\mathcal{M},\omega)$
an arbitrary representative, $ -
\frac{1}{k}\,i_{\vect{X}}(\omega^k)$ is a closed $(2k - 1)$-form.
Thus the de~Rham cohomology class of this form can be defined to
be $\pi_{2k-1}([\vect{X}]_{(2k-1)})$. It is easy to verify that
this definition is well defined: $\pi_{2k-1}([\vect{X}]_{(2k-1)})$
does not depend on the choice of the representative $\vect{X}$ in
$[\vect{X}]_{(2k-1)}$. As for the horizontal maps in the first row
of the above diagram, they are induced by the identity map on
$\XS^{(2k-1)}(M,\omega)=\XS^1(\mathcal{M},\omega)$ where $k \in
\{1,2,\ldots,n-1\}$. For example, if $[\vect{X}]_{(2k-1)}$ is an
equivalence class in
$H^{(2k-1)}_{\mathrm{EL}}(\mathcal{M},\omega)$ $( k=1, 2, \ldots,
n-1,\ n > 1 )$ where $\vect{X}\in\XS^{(2k-1)}(\mathcal{M},\omega)$
is an arbitrary representative, then $[\vect{X}]_{(2k-1)}$ is
mapped to be an equivalence class $[\vect{X}]_{2k+1}$ in
$H^{(2k+1)}_{\mathrm{EL}}(\mathcal{M},\omega)$. It is also easy to
check that this is a well defined homomorphism.

Since $\pi_1$ is an isomorphism and the horizontal homomorphisms
in the first row are all onto, it follows that \begin{thm} For $2
\leqslant k \leqslant n-1$, $\pi_{2k-1}$ is onto if and only if
$L_{\omega}^{k-1}=L_{\omega^{k-1}}$ from
$H^{(1)}_{\mathrm{dR}}(\mathcal{M})$ to
$H^{(2k-1)}_{\mathrm{dR}}(\mathcal{M})$ is onto, and
$L_{\omega^{k-1}} : \HdR^{(1)}(\mathcal{M}) \longrightarrow
\HdR^{(2k-1)}(\mathcal{M})$ is injective if and only if the
homomorphism from $\HEL^1(\mathcal{M},\omega)$ to
$\HEL^{(2k-1)}(\mathcal{M},\omega)$ is injective. \label{T:dec}
\end{thm}

\begin{corol} For $n \geqslant 3$, let
$\mathcal{M}$ be the torus $T^{2n}$ with the standard symplectic
structure $\omega$. Then, for $1 < k < n$,
$\HEL^{(2k-1)}(\mathcal{M},\omega) \neq \HdR^{(2k-1)}(\cal M)$.
\label{cor:T2n}
\end{corol}
\begin{proof} As is well known, the de~Rham cohomology groups
of $T^{2n}$ satisfy \begin{displaymath}
  \dim \HdR^{(k)}(T^{2n}) = {{2n}\choose{k}}
\end{displaymath}
for each $0 \leqslant k \leqslant 2n$. Therefore, we have
$\dim\HdR^{(2k-1)}(T^{2n}) > 2n$ for each $1 < k < n$. On the
other hand, due to the fact that the maps in the first row of the
diagram (\ref{eq:CD}) are surjective, we have
$\dim\HEL^{(2k-1)}(T^{2n},\omega) \leqslant 2n$ for each $1 < k <
n$. So, $\dim\HdR^{(2k-1)}(T^{2n})
>\dim\HEL^{(2k-1)}(T^{2n},\omega)$.
\end{proof}

Further, we will show  that there are some symplectic manifolds
for which $\HEL^{(2k-1)} \neq \HEL^{(1)}$.

Recall that on an $n$-dimensional Lie group $G$ there exists a
basis that consists of $n$ left-invariant vector fields
$\vect{X}_1$, \ldots, $\vect{X}_n$. They form the Lie algebra
$\mathfrak{g}$ of $G$. Let $[\vect{X}_i, \vect{X}_j] =-
c_{ij}^k\,\vect{X}_k$ with  the structural constants $c_{ij}^k$ of
$\mathfrak{g}$. Let $\{\theta^k\}$ be the left-invariant dual
basis. Then they satisfy the equation
\begin{equation}
  \dd\theta^k =\frac{1}{2}\,c_{ij}^k\,\theta^i \wedge\theta^j,
  \quad
k=1,\ldots, n.
\end{equation}

$G$ is called a \deff{nilpotent} Lie group if $\mathfrak{g}$ is
nilpotent. A {\it nilmanifold} is defined to be a closed manifold
$M$ of the form $G/\Gamma$ where $G$ is a simply connected
nilpotent group and $\Gamma$ is a discrete subgroup of $G$. It is
well known that $\Gamma$ determines $G$ and is determined by $G$
uniquely up to isomorphisms (provided that $\Gamma$ exists)
\cite{M,R}.

There are three important facts for the compact nilmanifolds
\cite{TO}:
\begin{enumerate}
\item Let $\mathfrak{g}$ be a nilpotent Lie algebra with
structural constants $c_{ij}^k$ with respect to some basis, and
let $\{\theta^1,\ldots ,\theta^n\}$ be the dual basis of
$\mathfrak{g}^{\ast}$. Then in the Chevalley-Eilenberg complex
$(\Lambda^{\ast}\mathfrak{g}^{\ast},\dd)$ we have
\begin{equation}\label{cte}
\dd\theta^k=\sum_{1\leqslant i<j<k} c_{ij}^k \,\theta^i\wedge
\theta^j, \quad k=1,\ldots, n.
\end{equation}
\item Let $\mathfrak{g}$ be the Lie algebra of a simply connected
nilpotent Lie group $G$. Then, by Malcev's theorem \cite{M}, $G$
admits a lattice if and only if $\mathfrak{g}$ admits a basis such
that all the structural constants are rational. \item By Nomizu's
theorem, the Chevalley-Eilenberg complex
$(\Lambda^{\ast}\mathfrak{g}^{\ast},\dd)$ of $\mathfrak{g}$ is
quasi-isomorphic to the de~Rham complex of $G/\Gamma$. In
particular,
\begin{equation}\label{nom}
H^{\ast}(G/\Gamma)\cong H^{\ast}(\Lambda^{\ast}\mathfrak{g}^*,\dd)
\end{equation}
and any cohomology class $[\alpha]\in H^k(G/\Gamma)$ contains a
homogeneous representative $\alpha$. Here we call the form
$\alpha$ homogeneous if the pullback of $\alpha$ to $G$ is
left-invariant.
\end{enumerate}

These results allow us to compute cohomology invariants of
nilmanifolds in terms of the Lie algebra $\mathfrak{g}$, and this
simplifies the calculations.

\begin{thm} There exists a $6$-dimensional
symplectic nilmanifold $(\mathcal{M}^6,\omega)$ such that
$$\HEL^{(3)}(\mathcal{M},\omega)\neq \HEL^{(1)}({\cal M},\omega).$$
\label{thm:nilm}
\end{thm}
\begin{proof} To define the manifold $\mathcal{M}$, it
suffices to give the Lie algebra. $\mathfrak{g}$ is a
$6$-dimensional Lie algebra generated by the generators
$\vect{X}_1,\ldots,\vect{X}_6$ with Lie bracket given by
$$[\vect{X}_i,\vect{X}_j]=-\sum_{1\leqslant i<j<k} c_{ij}^k
\vect{X}_k.$$ This Lie algebra gives a unique nilmanifold
$\mathcal{M}$ by the above information on nilmanifolds. The
Chevalley-Eilenberg complex
$(\Lambda^{\ast}\mathfrak{g}^{\ast},\dd)$ of $\mathfrak{g}$, which
is used to calculate the de~Rham cohomology of $\mathcal{M}$, is
as in the following.

Let $A=\Lambda^*(\theta^1,\ldots,\theta^6)$ be generated by the
1-forms $\theta^i, ~1\leqslant i \leqslant 6$. Their differentials
are given by the following formulas:
\begin{displaymath}
\begin{array}{lcl}
  \dd \theta^1 = 0, & & \dd \theta^2 = 0, \\
  \dd \theta^3 = 0, & &
  \dd \theta^4=\theta^1\wedge \theta^2 , \\
  \dd \theta^5=\theta^1\wedge \theta^4 -\theta^2\wedge \theta^3, & &
  \dd \theta^6=\theta^1\wedge \theta^5 +\theta^3\wedge \theta^4.
\end{array}
\end{displaymath}
Furthermore, the symplectic form $\omega$ on $\mathcal{M}$ is
induced by $F=\theta^1\wedge \theta^6 +\theta^2\wedge \theta^4
+\theta^3\wedge \theta^5 \in A$. $\omega$ is a symplectic form
since $F\wedge F\wedge F$ is nontrivial by an easy calculation.

It is not difficult to show
$$\HdR^{(1)}({\cal M}) = \mathbb{R}^3
  = \Span\{[\theta^1],[\theta^2],[\theta^3]\}$$
and $$\HdR^{(2)}({\cal M})=\mathbb{R}^4 = \Span\{[\theta^1\wedge \theta^3],
[\theta^1\wedge \theta^4], [\theta^2\wedge \theta^4], [F]\}.$$

To prove the theorem, it is sufficient to prove that $\omega
\wedge \theta^1$ is cohomologically trivial, according to
Theorem~\ref{T:dec}. This follows the equation

$$F \wedge \theta^1=\theta^1\wedge \theta^2\wedge \theta^4
+\theta^1\wedge \theta^3\wedge \theta^5 =\dd(\theta^2\wedge
\theta^5 +\theta^3\wedge \theta^6),$$ which can be checked easily
and thus completes the proof.
\end{proof}

\subsection{The Euler-Lagrange Cohomology and  the Harmonic Cohomology}

On a given symplectic manifold $(\mathcal{M},\omega)$, there also
exists the harmonic cohomology in addition to the de~Rham
cohomology.  In this subsection we explore the relation between
the Euler-Lagrange cohomology and the harmonic cohomology on
$(\mathcal{M},\omega)$, and show that they are different from each
other in general.

Given a smooth symplectic  manifold $(\mathcal{M},\omega)$, let
the $*$-operator
$$  *: \Omega^k(\mathcal{M}) \to
\Omega^{2n-k}(\mathcal{M})$$ be introduced \cite{Y} in analog with
the $*$-operator on a Riemannian manifold. Define
$$\delta: \Omega^k(\mathcal{M}) \to \Omega^{k-1}(\mathcal{M}), \quad
\delta(\alpha)=(-1)^{k+1}*\dd *\,\alpha. $$ It turns out to be
that $\delta=[i(\Pi),\dd]$ \cite{Y,Br}, where $i(\Pi)$ is, in
fact, the operator $\f$ introduced in subsection
\ref{sect:Operators}.

{\bf Remark 1:} The operator $\delta=-*\,\dd*$ was also considered
by Libermann (see \cite{LM}). Koszul~\cite{K} introduced the
operator $\delta=[\dd,i(\Pi)]$ for Poisson manifolds.
Brylinski~\cite{Br} proved that these operators coincide with each
other.

\begin{definition}
A form $\alpha$ on a symplectic manifold $(\mathcal{M},
\omega)$ is called {\it symplectically harmonic} if $d\alpha
=0=\delta \alpha$.
\end{definition}

We denote by $\Omega^k_{\mathrm{hr}}(M)$ the linear space of
symplectically harmonic $k$-forms.  Unlike the Hodge theory, there
are non-zero exact symplectically harmonic forms. Now, following
Brylinski~\cite{Br}, we define symplectically harmonic cohomology
$H^*_{\mathrm{hr}}(\mathcal{M},\omega)$ by setting
$$
H^k_{\mathrm{hr}}(\mathcal{M},\omega):=\Omega^k_{\mathrm{hr}}(\mathcal{M})/
(\im(\dd) \cap \Omega^k_{\mathrm{hr}}(\mathcal{M})).
$$
Therefore, $H^k_{\mathrm{hr}}(\mathcal{M},\omega)
\subset\HdR^k(\mathcal{M})$.

We would like to know if the symplectically harmonic cohomology
and the Euler-Lagrange cohomology are isomorphic to each other.
The following result answers no to this question.
\begin{thm} Let $\mathcal{M}$ be the $2n$-dimensional torus $T^{2n}$ with
standard symplectic structure. Then
$H^{(2k-1)}_{\mathrm{EL}}(\mathcal{M},\omega)$ and
$H^{(2k-1)}_{\mathrm{hr}}(\mathcal{M},\omega)$ are not the same
for $1<k<n$. \end{thm}

This is because in this case the symplectically harmonic
cohomology are the same as the de~Rham cohomology and now the
result follows from Corollary~\ref{cor:T2n}.

\subsection{The Relative Euler-Lagrange Cohomology}
\label{sect:RelativeEL}

Let us now propose a definition of relative Euler-Lagrange
cohomology. That is the combination of the above definition of
Euler-Lagrange cohomology and the usual definition of relative
de~Rham cohomology.

Let $\mathcal{M}$ be a symplectic $2n$-manifold and $i:
\mathcal{N} \longrightarrow \mathcal{M}$ be an embedded
sub\-manifold. Recall that the usual relative de~Rham forms are
defined by
  $$\Omega^k(i)=\Omega^k(\mathcal{M}) \oplus \Omega^{k-1}(\mathcal{N})$$
where $\Omega^{k-1}(\mathcal{N})$ is the group of $(k-1)$-forms on
$\mathcal{N}$. The differential on $\Omega^*(i)$ is given by
\begin{equation}
  \dd(\theta_1,\theta_2)=(\dd\theta_1,i^*\theta_1 - \dd\theta_2).
\end{equation}

\begin{definition}
Define
  $$\OmegaEL^{(2k-1)}(i,\omega)=\OmegaEL^{(2k-1)}(\mathcal{M},\omega)
  \oplus\Omega^{2k-2}(\mathcal{N})$$ where
  $\OmegaEL^{(2k-1)}(\mathcal{M},\omega)=\{i_{\vect{X}}(\omega ^k)\,|\, \vect{X}
  \in \mathcal{X}(\mathcal{M})\}$.
The \deff{relative Euler-Lagrange cohomology} will be defined as
$$\HEL^{(2k-1)}(i,\omega)=\frac{\{(\theta_1 , \theta_2)\in
 \OmegaEL^k(i,\omega) \,|
\ \dd(\theta_1 , \theta_2)=0 \}}{\{(\theta_1 , \theta_2)\,|\,
(\theta_1 , \theta_2)=\dd(\theta'_1 , \theta'_2)\}}.$$
\end{definition}

Let us consider an example for the relative Euler-Lagrange
cohomology. Let $\mathcal{M}=\R^{2n}$, $\mathcal{N}=T^n$ and
$i:T^n \longrightarrow \R^{2n}$ be the inclusion.
\begin{proposition}
 $\HEL^{(2k-1)}(i)=H_{\rm d R}^{(2k-2)}(T^n).$
\end{proposition}
\begin{proof}
There is the obvious linear map $\Omega^{2k-2}(T^n)
\longrightarrow \OmegaEL^{2k-1}(i)$ given by $\theta \longmapsto
(0,\theta)$ . When $\theta = \dd\alpha\in\Omega^{2k-2}(T^{2n})$ is
exact, $(0,\theta) = \dd(0,-\alpha)\in\OmegaEL^{2k-1}(i)$ is also
exact. Therefore a linear map $f:
\HdR^{(2k-2)}(T^{2n})\longrightarrow\HEL^{(2k-1)}(i),$ $[\theta]
\longmapsto [(0,\theta)]$ can be induced.

This map $f$ is an injection. In fact, if
$(0,\theta)=\dd(\alpha_1,\alpha_2)$, then $\dd \alpha_1=0$,
$\theta=i^*(\alpha_1)-\dd \alpha_2.$ Thus $\theta$ is exact by the
fact that any closed form $\alpha_1$ on $\R^{2n}$ is exact.

The map $f$ is also an epimorphism: For any closed
$(\theta_1,\theta_2)$, it is in the same cohomology class of the
element $(0, \theta_2-i^*(\alpha_1)+\dd \alpha_2) = (\theta_1,\theta_2)
-\dd(\alpha_1, \alpha_2)$, where $\theta_1=\dd
\alpha_1$ and $\alpha_2$ is any form on $T^n$. Obviously,
$\theta_2-i^*(\alpha_1)+\dd \alpha_2$ is closed and
$f([\theta_2-i^*(\alpha_1)+\dd \alpha_2]) = [(\theta_1,\theta_2)].$
\end{proof}

\vskip 2mm 
{\bf Remark 2:} Although it is not verified yet, the following
statement, if true, will not be a surprise : There exists a
symplectic manifold $\cal M$ and its submanifold $i:\mathcal{N}
\longrightarrow \mathcal{M}$ for which the (relative)
Euler-Lagrange cohomology is not the corresponding (relative)
de~Rham cohomology.

\vskip 2mm 
{\bf Remark 3:} For the definition of $\HEL^{(2k-1)}(i,\omega)$,
it is also possible to require that $(\theta'_1,\theta'_2)$ belong to
$\Omega^{2k-1}(i) = \Omega^{2k-1}(\mathcal{M})\oplus\Omega^{2k-2}(\mathcal{N})$
rather than $\OmegaEL^{2k-1}(i,\omega)$. The remaining explanations are
similar in principle.

\section{The Highest Euler-Lagrange Cohomology and
Volume-Preserving Systems }

Let us focus on the issues relevant to the volume-preserving
systems.

First,  it can be easily proved that, for each $k = 1, \ldots,
n-2$ and each point $x \in \mathcal{M}$, a 2-form
$\alpha\big|_x\in\Lambda_2(T^*_x\mathcal{M})$ satisfies
$\alpha\big|_x\wedge\omega^k = 0$ iff $\alpha\big|_x = 0$. As a
consequence, a smooth 2-form, $\alpha \in \Omega^2(\mathcal{M})$
satisfies $\alpha\wedge\omega^k = 0$ iff $\alpha = 0$ everywhere.
In other words, the linear maps
\begin{eqnarray*}
  \iota_k: \Omega^2(\mathcal{M}) & \longrightarrow & \Omega^{2k+2}(\mathcal{M})
\\
  \alpha & \longmapsto & \alpha\wedge\omega^k,\quad~1\leqslant k\leqslant n-2 ~.
\end{eqnarray*}
are injective. Thus we can obtain that
\begin{widetext}
\begin{displaymath}
  \XS^{(1)}(\mathcal{M},\omega) = \XS^{(3)}(\mathcal{M},\omega) = \ldots
  = \XS^{(2n-3)}(\mathcal{M},\omega) \subseteq \XS^{(2n-1)}(\mathcal{M},\omega).
\end{displaymath}
\end{widetext}

Especially, when $k = n-2$, the linear map
\begin{eqnarray}
  \iota = \iota_{n-2}: \Omega^2(\mathcal{M}) & \longrightarrow &
  \Omega^{2n-2}(\mathcal{M})
\nonumber \\
  \alpha & \longmapsto & \alpha\wedge\omega^{n-2}
\end{eqnarray}
is an isomorphism. If $n = 2$, we use the convention that
$\omega^0 = 1$, namely, $\iota = \mathrm{id}: \alpha \longmapsto
\alpha$. Then we can define a linear map $\phi$ making the
following diagram commutative:
\begin{equation}
  \begin{CD}
    \Omega^2(\mathcal{M}) @>\iota>> \Omega^{2n-2}(\mathcal{M}) \\
    @V\phi VV                        @V\dd VV\\
    \XH^{(2n-1)}(\mathcal{M},\omega) @>\nu_n>>
    B^{2n-1}(\mathcal{M}),
  \end{CD}
\label{phicg}
\end{equation}
where $\nu_n$ the linear isomorphism sending a vector field
$\vect{X}$ to a $(2n-1)$-form $- i_{\vect{X}}(\omega^n)$
\cite{gpwz,gpwz1}.

 It should be mentioned that, equivalently, given a 2-form
\begin{equation}
  \alpha = \frac{1}{2}\,Q_{ij}\,\dd q^i\wedge\dd q^j
  + A^i_j\,\dd p_i\wedge\dd q^j + \frac{1}{2}\,P^{ij}\,\dd p_i\wedge\dd p_j
\label{alpha}
\end{equation}
where $Q_{ij}$ and $P^{ij}$ satisfy
\begin{equation}
  Q_{ji} = - Q_{ij}, \qquad P^{ji} = - P^{ij},
\end{equation}
the vector field
\begin{displaymath}
  \phi(\alpha) = (\nu_n^{-1}\circ\dd\circ\iota)(\alpha)
  = \nu^{-1}_n(\dd\alpha\wedge\omega^{n-2})
\end{displaymath}
is in $\XH^{(2n-1)}(\mathcal{M},\omega)$.

For convenience, we set
\begin{displaymath}
  \vect{X} = n(n-1)\,\phi(\alpha)
  = n(n-1)\,\nu_n^{-1}(\dd\alpha\wedge\omega^{n-2}),
\end{displaymath}
namely, $i_{\vect{X}}(\omega^n) = -
n(n-1)\,(\dd\alpha)\wedge\omega^{n-2}$. It is easy to obtain that
\begin{equation}
  \vect{X} = \bigg(\frac{\partial P^{ij}}{\partial q^j}
  + \frac{\partial A^j_j}{\partial p_i} - \frac{\partial A^i_j}{\partial p_j}
  \bigg)\,\frac{\partial}{\partial q^i}
  + \bigg(\frac{\partial Q_{ij}}{\partial p_j}
  - \frac{\partial A^j_j}{\partial q^i}
  + \frac{\partial A^j_i}{\partial q^j}\bigg)\,
  \frac{\partial}{\partial p_i}.
\label{XH}
\end{equation}
 In fact, it can be easily obtained that
\begin{eqnarray*}
  \dd\iota(\alpha)&=&\dd\alpha\wedge\omega^{n-2}
\\  \omits{
  &=& \frac{1}{2}\,\frac{\partial Q_{ij}}{\partial q^k}\,
  \dd q^i\wedge\dd q^j\wedge\dd q^k\wedge\omega^{n-2}
  +\frac{1}{2}\,\frac{P^{ij}}{\partial p_k}\,
    \dd p_i\wedge\dd p_j\wedge\dd p_k\wedge\omega^{n-2}
\\
  & & +\bigg(\frac{1}{2}\,\frac{\partial Q_{jk}}{\partial p_i}
    + \frac{\partial A^i_j}{\partial q^k}\bigg)\,
    \dd p_i\wedge\dd q^j\wedge\dd q^k\wedge\omega^{n-2}
\\
  & & + \bigg( \frac{1}{2}\,\frac{\partial P^{ij}}{\partial q^k}
    -\frac{\partial A^i_k}{\partial p_j} \bigg)\,
    \dd p_i\wedge\dd p_j\wedge\dd q^k\wedge\omega^{n-2}
\\ }
  &=&  \bigg(\frac{1}{2}\,\frac{\partial Q_{jk}}{\partial p_i}
    + \frac{\partial A^i_j}{\partial q^k}\bigg)\,
    \dd p_i\wedge\dd q^j\wedge\dd q^k\wedge\omega^{n-2}
\\
  & & + \bigg( \frac{1}{2}\,\frac{\partial P^{ij}}{\partial q^k}
    -\frac{\partial A^i_k}{\partial p_j} \bigg)\,
    \dd p_i\wedge\dd p_j\wedge\dd q^k\wedge\omega^{n-2}.
\end{eqnarray*}
By virtue of the following two equations
\begin{widetext}
\begin{eqnarray*}
  \dd p_i\wedge\dd p_j\wedge\dd q^k \wedge\omega^{n-2}
  & = & \frac{\delta^k_j}{n-1}\,\dd p_i \wedge\omega^{n-1}
  - \frac{\delta^k_i}{n-1}\,\dd p_j \wedge\omega^{n-1},
\\
  \dd p_i\wedge\dd q^j\wedge\dd q^k\wedge\omega^{n-2}
  & = & \frac{\delta^j_i}{n-1}\,\dd q^k\wedge\omega^{n-1}
  - \frac{\delta^k_i}{n-1}\,\dd q^j\wedge\omega^{n-1},
\end{eqnarray*}
\end{widetext}
we can write $\dd\iota(\alpha)$ as the inner product of certain a
vector with $\omega^n$. \omits{
\begin{eqnarray*}
  \dd\iota(\alpha)
  & = & \frac{1}{n-1}\bigg(\frac{\partial A^j_j}{\partial q^i}
  - \frac{\partial A^j_i}{\partial q^j} - \frac{\partial Q_{ij}}{\partial p_j}
  \bigg)\,\dd q^i\wedge\omega^{n-1}
\nonumber \\ & &
  + \frac{1}{n-1}\bigg(\frac{\partial P^{ij}}{\partial q^j}
  + \frac{\partial A^j_j}{\partial p_i} - \frac{\partial A^i_j}{\partial p_j}
  \bigg)\,\dd p_i\wedge\omega^{n-1}
\\
  & = & \frac{1}{n(n-1)}\bigg(\frac{\partial A^j_j}{\partial q^i}
  - \frac{\partial A^j_i}{\partial q^j} - \frac{\partial Q_{ij}}{\partial p_j}
  \bigg)\,i_{\frac{\partial}{\partial p_i}}\omega^n
\nonumber \\ & &
  - \frac{1}{n(n-1)}\bigg(\frac{\partial P^{ij}}{\partial q^j}
  + \frac{\partial A^j_j}{\partial p_i} - \frac{\partial A^i_j}{\partial p_j}
  \bigg)\,i_{\frac{\partial}{\partial q^i}}\omega^n.
\end{eqnarray*}
} Comparing it with
\begin{equation}
  \dd\iota(\alpha) = \frac{1}{n(n-1)}\,\nu_n(\vect{X})
  = -\frac{1}{n(n-1)}\,i_{\vect{X}}(\omega^n),
\label{nunX}
\end{equation}
we can obtain the expression of $\vect{X}$, as shown in
eq.~(\ref{XH}).

Since both $\iota$ and $\nu_n$ are linear isomorphisms, we can see
from the commutative diagram (\ref{phicg}) that, for each 2-form
$\alpha$ on $\mathcal{M}$ as in eq.~(\ref{alpha}), the vector
field $\vect{X}$ in eq.~(\ref{XH}) belongs to
$\XH^{(2n-1)}(\mathcal{M},\omega)$; Also, for each
  $\vect{X}\in\XH^{(2n-1)}(\mathcal{M},\omega)$
there exists the 2-form $\alpha$ on $\mathcal{M}$ satisfying
eq.~(\ref{XH}). But there may be several 2-forms that are mapped
to the same vector field $\vect{X}$. For example, the vector field
$\vect{X}$ in (\ref{XH}) is invariant under the transformation
\begin{equation}
  \alpha \longmapsto \alpha + \theta
\label{transf}
\end{equation}
where $\theta$ is a closed 2-form.

Note that for the vector field
$\vect{X}\in\XH^{(2n-1)}(\mathcal{M},\omega)$, the 2-form $\alpha$
is a globally defined. If
$\vect{X}\in\XS^{(2n-1)}(\mathcal{M},\omega)$, such a 2-form
cannot globally be given if
$H^{(2n-1)}_{\mathrm{dR}}(\mathcal{M})$ is nontrivial. In this
case, $\alpha$ can be still found as a locally defined 2-form.
Then the relation between $\vect{X}$ and the locally defined
2-form $\alpha$, eq.~(\ref{XH}), is valid only on an  open subset
of $\mathcal{M}$.

It is clear that no matter whether $\vect{X}$ belongs to
$\XH^{(2n-1)}(\mathcal{M},\omega)$ or
$\XS^{(2n-1)}(\mathcal{M},\omega)$, i.e., whether the 2-form
$\alpha$ is globally or locally defined, the flow of $\vect{X}$
can be always obtained provided that the general solution of the
following equations can be obtained:
\begin{eqnarray}
  \dot{q}^{\,i} & = & \frac{\partial P^{ij}}{\partial q^j}
  + \frac{\partial A^j_j}{\partial p_i}
  - \frac{\partial A^i_j}{\partial p_j},
\nonumber \\
  \dot{p}_i & = & \frac{\partial Q_{ij}}{\partial p_j}
  - \frac{\partial A^j_j}{\partial q^i}
  + \frac{\partial A^j_i}{\partial q^j}.
\label{evps}
\end{eqnarray}
This is just the general form of the equations of a
volume-preserving mechanical system on a symplectic manifold
$(\mathcal{M},\omega)$.

\omits{ \vskip 4mm
\subsection{The Volume-Preserving Mechanical Systems}
 Now, let us illustrate some cases for the volume-preserving
systems.}

It should be pointed out that as volume-preserving systems the
ordinary Hamiltonian systems are included. \omits{Then we study
some non-Hamiltonian linear systems without potential but they
preserve the volume of the phase space. Finally, we emphasize the
importance of these volume-preserving systems.

For a function $H$ on $\mathcal{M}$, the Hamiltonian vector field
$\vect{X}_H$ preserves not only the volume-form, but also the
symplectic form $\omega$. Because of the relations in
(\ref{eq:Hrels}),} There should be a 2-form for the Hamiltonian
system, in fact, one of such a 2-form can be selected as
\begin{equation}
  \alpha = \frac{1}{n-1}\,H\,\omega.
  \label{eq:Hamilt}
\end{equation}
Substituting this 2-form into eqs.~(\ref{evps}), they turn to be
the ordinary canonical equations with $H$ as the
Hamiltonian.

Let us now consider the non-Hamiltonian linear system without
potential:
\begin{equation}
  \ddot{q}^{\ i} = - k_{ij} \,q^j
\label{ls}
\end{equation}
on $\mathbb{R}^{n}$ with constant coefficients $k_{ij}$ if
$k_{ij}\neq k_{ji}$.

It is obvious that eqs. (\ref{ls}) can be turned into the form:
\begin{equation}
  \dot{q}^{\ i} = \frac{\partial H}{\partial p_i}, \qquad
  \dot{p}_i = - \frac{\partial H}{\partial q^i} - a_{ij}\, q^j
\label{lsys}
\end{equation}
with
\begin{eqnarray}\nonumber
  & & H = \frac{1}{2}\,\delta^{ij}\,p_i p_j
  + \frac{1}{2}\, s_{ij}\,q^i q^j,
\label{lsysH} \\
  & & s_{ij} = \frac{1}{2}\,( k_{ij} + k_{ji}), \qquad
  a_{ij} = \frac{1}{2}\,(k_{ij} - k_{ji}).
\label{lsyscoef}
\end{eqnarray}
When one of $a_{ij}$ is nonzero, the system (\ref{lsys}) is just a
non-Hamiltonian system with non-potential force on
$\mathbb{R}^{2n}$. In fact, the corresponding vector field
\begin{displaymath}
  \vect{X} = \vect{X}_H - a_{ij}\,q^j\,\frac{\partial}{\partial p_i}
\end{displaymath}
of (\ref{lsys}) is not even a symplectic vector field since the
Lie derivative with respect to this vector field does not preserve 
the symplectic form. \omits{
\begin{displaymath}
  \Lied{X}\omega = a_{ij}\,\dd q^i\wedge\dd q^j
\end{displaymath}
does not vanish in this case. On the other hand,}However, the
linear system (\ref{lsys}) always preserves the volume form of the
phase space because $\Lied{X}(\omega^n) = 0$.

If we take
\begin{equation}
  \alpha = \frac{1}{n-1}\,H\,\omega
  - \frac{1}{2}\,p_k q^k\,a_{ij}\,
  \dd q^i\wedge\dd q^j,
\label{lsysalpha}
\end{equation}
then the general equations (\ref{evps})  turn into
eqs.~(\ref{lsys}). Therefore the 2-form $\alpha$ in
eq.~(\ref{lsysalpha}) is the general 2-form \omits{up to an exact
form} corresponding to a linear system without potential.

It should be mentioned that 
even a conservative system can be transformed to be a
non-Hamiltonian linear system. For example, two 1-dimensional
linearly coupled oscillators may construct such a system:
\begin{displaymath}
  m_1\,\ddot{q}^{\,1} = -k\,(q^2 - q^1), \qquad
  m_2\,\ddot{q}^{\,2} = -k\,(q^1 - q^2).
\end{displaymath}
Obviously, such a system satisfies Newton's laws. But, it is not a
Hamiltonian system if $m_1\neq m_2$: Let $k_{11} = -k/m_1$,
$k_{12} = k/m_1$, $k_{21} = k/m_2$ and $k_{22} = -k/m_2$. Then it
is a system as described by eqs.~(\ref{ls}). When $m_1 \neq m_2$,
we have a system satisfying $k_{12} \neq k_{21}$.

\section{Liouville's Theorem and Its Generalizations}

\subsection {$H^{(1)}_{EL}({\cal M})$ and Symplectic Liouville's Theorem}
\vskip 0mm

 It is well known that for a given  differentiable vector field
$X$ on $\mathcal{M}$, the corresponding flow is defined by\omits{ Given a
differentiable vector field $X \in T{\cal M}$,} the 1-parameter
transformation group on $({\cal M}, \omega)$
\begin{eqnarray*}
\varphi^t: {\cal M}&\rightarrow & {\cal M}, \quad
p\mapsto q=\varphi^t(p), \\
\varphi _*^{t}:T{\cal M}&\rightarrow &T{\cal M}, \quad X_p\mapsto
X_q=\varphi _*^{t}X_p
\end{eqnarray*}
such that $\varphi _t(p)$ is the integral curve of $X$
 from  $p \in {\cal M}$. The 1-parameter transformation group
$\{\varphi^t\} \in Diff({\cal M})$ is called the flow with respect to
$X$.

The {\it phase flow} is defined by the 1-parameter
transformation group on 
$({\cal M}, \omega)$
$$
g^t: (p_i(0), q^j(0))\mapsto(p_i(t), q^j(t)), \quad \forall
t\in[0,1],
$$
where $p_i(t), q^j(t)$ are solutions of canonical equations.
Namely, $X \in {\cal X}_H({\cal M},\omega)$, $g^t \in
Diff_H({\cal M})$. 

The famous {\bf Liouville's theorem} states\cite{Arnold}: The
phase flow preserves volume. Namely, for any region $D\subset
{\cal M}$,\omits{ \footnote{V.I. Arnold, Mathematical Methods of
CM, 2nd Ed. Springer-Verlag, (1989).}}

\centerline{\it volume of $g^t D$=volume of $ D$.} \omits{\vskip
0mm \qquad\qquad \qquad\qquad \qquad \qquad  {\small V.I. Arnold,
Mathematical Methods of CM (1989).}}
In order to distinguish with its generalizations via the
Euler-Lagrange cohomology groups, we call it 
the {\bf Hamiltonian Liouville's theorem}, which is corresponding
to the image of the first Euler-Lagrange cohomology group.

However, it is also well known that the symplectic flow generated
by a symplectic vector field preserves the volume as well
\cite{lecture} and the symplectic vector field is in the kernel of
the first Euler-Lagrange cohomology group $H^{(1)}_{EL}({\cal
M},\omega)$. Eventually, the \omits{definition for {\it symplectic
flow} reads: The }1-parameter
transformation group on 
$({\cal M}, \omega)$, $\{f^t\} \in Diff_S({\cal M}, \omega)$, is
called a {\it symplectic flow}. If
\be\label{symflow}
 f^t_*: X \mapsto X_t:=f^t_* X, \quad X_{t=0}=X,~t \in [0, 1] %
\ee%
where $X_t\in {\cal X}^{(1)}_S({\cal M},\omega), \forall t \in [0, 1]$,
i.e. they are {\it symplectic vectors.}
Thus, we may generalize famous Liouville's theorem to its
symplectic counterpart.

\begin{thm}[Symplectic Liouville's Theorem] The symplectic flow
preserves volume. Namely, for any region $D\subset \cal M$,
\begin{center}\it volume of $f^t D$=volume of $ D$.\end{center}
\end{thm}
This states that the necessary and sufficient
 condition for the conservation law of volume of $D \subset \Gamma$
 under symplectic flow $f^t$ is the Euler-Lagrange 1-form on $D$ is closed.


It is easy to prove this theorem. Since  $$v(t):=\int_{D(t)}\tau,
\qquad
D(t)=f^t D(0),
$$
where $\tau=\frac{1}{n!}\omega^n\omits{\underbrace{\omega \wedge
\cdots\wedge \omega }_{n}}$ is the volume element. Since
$$
\int_{D(t)}\tau=\int_{f^t D(0)}\tau=\int_{D(0)}f^{t *}\tau,
$$
and for the symplectic map $f^t$, by definition we have $ f^{t
*}\omega=\omega$. This leads to \omits{ ~\Rightarrow~}$f^{t
*}\tau=\tau$, from which it immediately follows  what we want to
get: $v(t)=v(0): ~\int_{D(t)}\tau=\int_{D(0)}\tau$. 
Thus, the proof is completed.

\omits{It should be noted the following points. First,  in
general, for $n \geq 2$ it is easy to construct volume-preserving
diffeomorphisms of ${R}^{2n}$ which are not
symplectic, and hence $
{Symp}({R}^{2n})$ is a proper subgroup of
$
{Diff}_{\mathrm{vol}} ({R}^{2n})$\cite{lecture}\omits{\footnote{D.
McDuff et. al. p21).}}. In fact, $ f_v^{t *}\tau=\tau$ requires
${\cal L}_{X_v}\omega=\varpi$, where $f_v^{t}
\in{Diff}_{\mathrm{vol}} ({R}^{2n}),~~{X_v}\in {\cal
X}_V({R}^{2n})$, such that
$\varpi\wedge \omega
=0$. Thus, ${\cal L}_{X_v}\tau=0.$
 Secondly, both (Hamiltonian) Liouville's
theorem and \omits{extended}symplectic one are sufficient for the
volume conservation but not necessary. Thus, the couple of
Liouville's theorems via the first cohomology for the phase volume
preserving property should be further generalized.}

\omits{Thirdly, it is well known that Hamiltonian Liouville's
theorem has particularly important applications in statistical
physics and allows to apply methods of ergodic theory to the study
of classical mechanics. The generalizations of Hamiltonian
Liouville's theorem should be made step by step in order that
these applications could also be possible.}

\subsection{$H^{(2k-1)}_{EL}({\cal M},\omega)$ and Generalized Liouville's
  Theorem}

Let us now consider how to generalize the couple of the symplectic
and Hamiltonian Liouville's theorems further.

As was noted, the symplectic and Hamiltonian vector fields
generating the symplectic and phase flows correspond to the kernel
and image of the first Euler-Lagrange cohomology, respectively.
Therefore,  it is reasonable to generalize these theorems
according to the higher Euler-Lagrange cohomology groups
$H^{(2k-1)}_{EL}({\cal M},\omega), 1<k \leqslant n$. For the kernel and
image of each of them, there should be a couple of the
$(2k-1)$-degree symplectic(-like) and Hamiltonian(-like) Liouville's
theorems, which claim that the $(2k-1)$-degree symplectic(-like) and
Hamiltonian(-like) flows, $f^{t(2k-1)}_S$ and $f^{t(2k-1)}_H$
generated by the $(2k-1)$-degree symplectic(-like) and
Hamiltonian(-like) vector fields, $\XS^{(2k-1)}$ and $\XH^{(2k-1)}$,
preserve the volume, respectively. For the highest one, the
$(2n-1)$st Euler-Lagrange cohomology group $H^{(2n-1)}_{EL}({\cal
M},\omega)$, its kernel and image characterize directly the
symplectic(-like) and Hamiltonian(-like) volume-preserving vector
fields, respectively. Consequently, the Hamiltonian(-like)
volume-preserving Liouville's theorem holds only if the general
equations of volume-preserving being satisfied, while the
symplectic(-like) volume-preserving Liouville's theorem holds  if
and only if  the flows $f^{t (2n-1)}_S$ are generated by the
symplectic(-like) volume-preserving vector fields $\XS^{(2n-1)}$ and
the general equations of volume-preserving being only locally
satisfied  if this cohomology group is not trivial.

For a given $k$, $1 \leqslant k < n$, let us first consider a
symplectic(-like) vector field
$\vect{X}\in\XS^{(2k-1)}(\mathcal{M},\omega)$. Assume that its
flow is ${f}^{t (2k-1)}_S$. As stated previously, $\vect{X}$ is a
symplectic vector field, in fact. Therefore we have
$\Lied{X}\omega = 0$, $\Lied{X}\omega^2 = 0$, \ldots,
$\Lied{X}\omega^n = 0$. By the definition of the Lie derivatives,
we have
\begin{widetext}
\begin{equation}
  ({f}^{t (2k-1)}_S)^*\omega = \omega, \quad
  ({f}^{t (2k-1)}_S)^*\omega^2 = \omega^2, \quad \ldots,
  ({f}^{t (2k-1)}_S)^*\omega^n = \omega^n
\end{equation}
\end{widetext}
for each possible parameter $t$. For a Hamiltonian(-like) vector
field $\vect{X}\in\XH^{(2k-1)}(\mathcal{M},\omega)$, it is almost
the same except the flow ${f}^{t (2k-1)}_S$ should be replaced by
${f}^{t (2k-1)}_H$.

Now consider a $2l$-dimensional orientable submanifold $i:
\sigma^{2l} \hookrightarrow\mathcal{M}$ where $1 < 2l \leqslant 2n$. It
is possible that $i^*\omega^l = 0.$ For example, when $2l \leqslant n$
and $\sigma^{2l}$ lies in a Lagrangian submanifold, the $2l$-form
$i^*\omega^l$ is zero. When it is nonzero,
$\frac{1}{l!}\,i^*\omega^l$ can be treated as a volume form on
$\sigma^{2l}.$ In this case the volume of the submanifold is
$\frac{1}{l!}\int_{\sigma^{2l}}\omega^l.$ Let ${f}^{t
(2k-1)}_L(\sigma^{2l}),~L=S,H,$ be the image of $\sigma^{2l}$
mapped by the corresponding flow ${f}^{t (2k-1)}_L$, respectively.
Then we have
\begin{widetext}
\begin{eqnarray}\label{areapreserv's}
  \frac{1}{l!}\int_{{f}^{t (2k-1)}_L(\sigma^{2l})}\omega^l
  = \frac{1}{l!}\int_{\sigma^{2l}}({f}^{t (2k-1)}_L)^*\omega^l
  = \frac{1}{l!}\int_{\sigma^{2l}}\omega^l,~~L=S,H.
\end{eqnarray}
\end{widetext}
That is, the volume of the submanifold $\sigma^{2l}$ will be
preserved when it is sent to another one along the flow ${f}^{t
(2k-1)}_{S, H}$ of a symplectic or Hamiltonian(-like) vector filed,
respectively.

However, for a symplectic(-like), including the Hamiltonian(-like),
volume-preserving vector field $\vect{X}\in
\XS^{(2n-1)}(\mathcal{M},\omega),
\XH^{(2n-1)}(\mathcal{M},\omega)$, the forms $\omega$, $\omega^2$,
\dots, $\omega^{n-1}$ are not necessarily preserved. In this case
we can consider only a region $D\subset \mathcal{M}$. It can be
similarly discussed that
\begin{eqnarray}\label{vp}
  \frac{1}{n!} \int_{{f}^{t (2n-1)}_{L}(D)}\omega^n
  = \frac{1}{n!}\int_{D}\omega^n,~~L=S,H.
\end{eqnarray}

In conclusion, the sequence of area preserving laws
(\ref{areapreserv'}) in classical mechanics should also be
generalized to a couple of sequences of the area preserving laws
(\ref{areapreserv's}), (\ref{vp}) according to
the Euler-Lagrange cohomological groups.\omits{ as follows%
\be\label{areapreserv's} \nonumber\int_{f^{t (1)}_L
\sigma^2}\omega &=& \int_{\sigma^2}\omega, ~\forall 2{\rm -chain}~
\sigma^2 \subset
{\mathcal{M}^{2n}},\\\nonumber%
&& \cdots\\\nonumber%
 \int_{f^{t (2k-1)}_L \sigma^{2k}}\omega^k &=&
\int_{\sigma^{2k}}\omega^k, ~\forall 2k{\rm -chain}~ \sigma^{2k}
\subset
{\mathcal{M}^{2n}},~1<k<n,\\
&& \cdots\\\nonumber
 \int_{f^{t (2n-1)}_L \sigma^{2n}}\omega^{n} &=& \int_{ \sigma^{2n}}\omega^{2n},
~\forall 2n{\rm -chain} ~\sigma^{2n} \subset
{\mathcal{M}^{2n}},\\\nonumber \ee%
where $L=S,H$.} Thus, we have

\begin{thm}[Generalized Liouville's Theorem] Given a vector field
$\vect{X}_L\in \XS^{(2k-1)}(\mathcal{M},\omega)$, $(L=S,H)$ on
$(\mathcal{M},\omega)$, its flow $f_{L}^{t (2k-1)}$, $(L=S,H)$,
preserves the volume. Namely, for any region $D\subset \cal M$,
\begin{center}\it volume of $f_{L}^{t
(2k-1)} D$=volume of $ D, \quad L=S,H$.\end{center}
\end{thm}%

As was mentioned, for the Hamiltonian(-like) sequence $(L=H)$ of the 
conservation laws, the corresponding equations of motion hold.
Namely, this sequence of conservation laws are ``on shell". For
the symplectic(-like) sequence $(L=S)$ of these conservation laws with $k < n$,
the relevant Hamiltonian in the equations of motion are defined locally if the
Euler-Lagrange cohomology group $\HEL^{(1)}(\mathcal{M},\omega)$ is not trivial.
For the symplectic(-like) volume systems $(L =S \textrm{ and } k = n)$, there
will
be no Hamiltonians (even defined locally). Instead, they are described by some
2-forms defined patch by patch provided that $\HEL^{(2n-1)}(\mathcal{M},\omega)$
is nontrivial.

It should be mentioned that this also generalizes the
famous Noether's theorem in classical mechanics. For a given
Hamiltonian system with symmetries, Noether's theorem always
relates all conservation laws with  equations of motion of the
system since the action is invariant under the symmetries.
\omits{\red
>>>}For the
symplictic(-like) Liouville's theorems, however, they relate the
corresponding simplicity conservation laws directly with the
kernel of the relevant Euler-Lagrange cohomology rather than
equations of motion of the original mechanical systems.

\section{Concluding Remarks}

In this paper, we have  introduced the definition of the
Euler-Lagrange cohomology groups
$\HEL^{(2k-1)}(\mathcal{M},\omega)$, $1\leqslant k \leqslant n,$
on symplectic manifolds $(\mathcal{M}^{2n},\omega)$ and studied
their relations with other cohomologies as well as some of their
properties \cite{zgpw}, \cite{zgw}, \cite{gpwz1}. It is shown
that, for $k=1, n $, $\HEL^{(1)}(\mathcal{M},\omega)$ and
$\HEL^{(2n-1)}(\mathcal{M},\omega)$ are isomorphic to the de~Rham
cohomology $\HdR^{(1)}(\mathcal{M})$ and
$\HdR^{(2n-1)}(\mathcal{M})$, respectively. On the other hand,
generally $\HEL^{(2k-1)}(\mathcal{M},\omega)$, $1< k < n,$ is
isomorphic neither to the de~Rham cohomology
$\HdR^{(2k-1)}(\mathcal{M})$ nor to the harmonic cohomology on
$(\mathcal{M}^{2n}, \omega)$. And they are also different from
each other in general. To our knowledge, these Euler-Lagrange
cohomology groups on $(\mathcal{M}^{2n}, \omega)$ have not yet
been introduced systematically before. It is significant to know
whether there are some more important roles played by these
cohomology groups to the symplectic manifolds.

The ordinary canonical equations in Hamilton mechanics correspond
to 1-forms that represent trivial element in the first
Euler-Lagrange cohomology $\HEL^{(1)}(\mathcal{M},\omega)$ on the
phase space. Analog to this property, the general
volume-preserving Hamiltonian equations on phase space are presented, from the
cohomological point of view, in terms of forms which represent the
trivial element in the highest Euler-Lagrange cohomology group
$\HEL^{(2n-1)}(\mathcal{M},\omega)$. And the ordinary canonical
equations in Hamilton mechanics become their special cases. It is
of course interesting to see what about the representatives in the
trivial elements of other Euler-Lagrange cohomology groups and
whether they also lead to some meaningful dynamical equations.
These problems are under investigation.

We have also introduced the conception of relative Euler-Lagrange
cohomology. It is of course interesting to see its applications in
mechanics and physics.

The first Euler-Lagrange cohomology group  has been introduced in
order to consider its time-discrete version in the study on the
discrete mechanics including the symplectic algorithm
\cite{ELcoh2,ELcoh4}. Although the first Euler-Lagrange cohomology
is isomorphic to the first de~Rham cohomology, its time discrete
version is still intriguing  and plays an important role in the
symplectic algorithm. In addition, it has also been introduced in
the field theory and their discrete versions of independent
variables. The latter is closely related to the multi-symplectic
algorithm \cite{ELcoh2,ELcoh4}. It is of course significant to
introduce the higher Euler-Lagrange cohomology groups in these
fields and to explore their applications. Furthermore, since the
general volume-preserving equations have been introduced on
symplectic manifold from the cohomological point of view, it is
meaningful to investigate their time-discrete version and study
its relation with the volume-preserving algorithm.

Based on the Euler-Lagrange cohomology groups and
 Hamiltonian(-like) volume-preserving equations on  symplectic
manifold $(\mathcal{M}, \omega)$, we have generalized famous
Liouville's theorem in classical mechanics. It has been first
generalized \omits{from the ordinary one with the phase flows of
the canonical equations }to the symplectic flows generated by the
symplectic vector fields, the kernel of the first Euler-Lagrange
cohomology group, 
i.e. to {\it symplectic Liouville's theorem}. Then, we have
further generalized the  symplectic and Hamiltonian Liouville's
theorems to the higher ones corresponding to  the kernel and image
of each higher Euler-Lagrange cohomology group $H^{(2k-1)}_{\rm
EL}(\mathcal{M}),~1<k\leqslant n$ on $(\mathcal{M}, \omega)$,
respectively, i.e.\omits{ Namely, we have established the couple
sequences of Liouville's theorems that correspond to the
$(2k-1)$st symplectic(-like) and Hamiltonian(-like)\omits{
volume-preserving} vector fields, respectively. The former
generate} the symplectic(-like)\omits{ volume-preserving flows $f^{t
(2k-1)}_S$ that do not require the equations of motion
globally\omits{ (\ref{evps})} and the latter generate the} and
Hamiltonian(-like) \omits{volume-preserving flows $f^{t (2k-1)}_H$
that do require the equations of motion\omits{ (\ref{evps})}.
\omits{ It should be worthwhile  pointing  out that first, for
each of the other Euler-Lagrange cohomology groups $H^{2k-1}_{\rm
EL}(\mathcal{M},\omega), 1<k<n$ on $(\mathcal{M}, \omega)$, there
should also exist a couple of the $2k-1$st symplectic(-like) and
Hamiltonian(-like)}} Liouville's theorems.\omits{ that can be
established by an analog manner presented in this paper.}

In all these 
Liouville's theorems, the symplectic(-like) (for $1\leqslant k \leqslant n$)
and the Hamiltonian(-like) (for $1<k \leqslant n$) ones have offered a
kind of generalizations of the famous Noether's
theorem, which 
does not require the systems to possess actions necessarily
in ordinary sense. In other words, the conservation laws may hold
with respect to the relevant cohomology and for the classical
systems with non-potential forces.

It should be mentioned that this kind of generalization of
Noether's theorem may also work for other cohomology groups of the
symmetries of the systems on the relevant manifolds. \omits{The
key point is that once there is a conservation law with respect to
the symmetry of the system via Noether's theorem, which always
requires that the equation of motion of the system should hold,
one should consider the cohomological aspects of the symmetry and
explore if the conservative property could be related to the
cohomology of the symmetry and the condition for the conserved
properties might be relaxed from the equation of motion. This may
also generalize to the case of field theory, where infinite
dimensional symmetries might be involved. In addition, for a given
conservation law with respect to the symmetry of the system, the
condition for the corresponding ``differential" conservation law
could also be weakened than the original one, i.e. the equation of
motion of the system  might no longer hold and it might be
replaced by some weaker condition. For example, it is known that
for the  stationary axial symmetric gravitational fields in some
gravity theory with diffeomorphism invariance, the mass formula
for the black holes with the stationary axial symmetry is a
vanishing Noether's charge with respect to the axial symmetry and
its ``differential" form is the fist law of the black hole
thermodynamics, for which the holding condition is weakened than
the equation of motion of the system \cite{blackhole}.} We will
explore these aspects in detail elsewhere.

Finally, it should also be pointed out that these 
Liouville's theorems should play important roles not only in classical and
continuous  mechanics, field theories 
but also especially in statistical physics.\omits{ as was
mentioned in the introduction and context. However, this has not
been noticed widely yet. \omits{In fact, famous Liouville's
theorem in mechanics concerns with the volume preserving processes
along the phase flow and plays a very important role in
statistical physics.}} In fact,\omits{ the distribution function
of probability is independent of time named} the statistical
Liouville's theorem, based on  Liouville's theorem in mechanics,
 is one of the fundamental principles in statistical physics.
According to our results, however, the generalized Liouville's
theorems\omits{ in mechanics should be generalized first from the
phase flow to the symplectic flow and further to the flow that
generated by the volume-preserving vector fields. Thus,} should
offer 
a way to generalize Liouville's theorem in statistical
physics.\omits{ should also be generalized.} We should also leave
this issue for the forthcoming publication.

\begin{acknowledgments}The authors would like to thank Professors
C.-G. Huang, Z.J. Shang, X.Y. Su, S.K. Wang and K. Wu for valuable
discussions. This work is partly supported by NSFC (Nos. 90103004,
10175070, 10375087).

\end{acknowledgments}

\omits{\section*{Appendix: }}

\end{document}